\documentclass[preprint,fleqn,prb,onecolumn,12pt]{revtex4}%
\usepackage{amssymb}
\usepackage{amsmath}
\usepackage{graphicx}%
\usepackage{amsfonts}
\begin{document}
\title{ Nondiffractive sonic crystals}
\author{Isabel P\'{e}rez-Arjona$^{1}$, V\'{\i}ctor J. S\'{a}nchez-Morcillo$^{1}$,
Javier Redondo$^{1}$, V\'{\i}ctor Espinosa$^{1}$ and Kestutis Staliunas$^{2}$}
\affiliation{$^{1}$Departamento de F\'{\i}sica Aplicada, Escuela Polit\'{e}cnica Superior
de Gandia, Universidad Polit\'{e}cnica de Valencia, Crta. Natzaret-Oliva s/n,
46730 Grau de Gandia, Spain.}
\affiliation{$^{2}$ICREA, Departament de Fisica i Enginyeria Nuclear, Universitat
Polit\`{e}cnica de Catalunya, Colom 11, E-08222 Terrassa, Barcelona, Spain}
\author{Short title: Nondiffractive sonic crystals}
\maketitle

\pagebreak

\begin{center}
ABSTRACT
\end{center}

We predict theoretically the nondiffractive propagation of sonic waves in
periodic acoustic media (sonic crystals), by expansion into a set of plane
waves (Bloch mode expansion), and by finite difference time domain
calculations of finite beams. We also give analytical evaluations of the
parameters for nondiffractive propagation, as well as the minimum size of the
nondiffractively propagating acoustic beams.

\vspace{1.5cm}

PACS numbers: 43.20.Fn, 43.35.Cg, 43.20.Wd

\vspace{15cm}

\section{Introduction}

The study of the dynamics of waves, even in simple linear media, initiated
hundreds of year ago, ever and ever leads to surprisingly new results and
insights. One of such \textquotedblleft surprises\textquotedblright\ was the
discovery of band gaps in the propagation of light in materials with the
refraction index periodically modulated on the scale of the optical
wavelength, the so called photonic crystals$^{1}$. The theory of wave
propagation in periodic materials was developed long time ago by Bloch and
Floquet, and it found many applications in solid state physics, in particular
in the studies of electronic properties of semiconductors (calculation of
valence and conduction bands, etc). Nevertheless, the advent of the photonic
crystals initiated a revival of the theory of wave propagation in periodic
media. The creation and control of photonic band gaps$^{2}$, the slowing down
of light$^{3}$, and the photonic crystal waveguides are the main applications
to the date. Most of these studies concern the propagation of plane waves (not
the beams), and results in the modification of the\textit{\ temporal
}dispersion relation (frequency versus propagation wavenumber). Later, the
strong analogies between the propagation of light and sound (which obey
similar wave equations) motivated the study of sound propagation in periodic
acoustic media, the so called sonic or phononic crystals (SC). Many of the
results obtained in the photonic case have been reported in the sonic case.
For a review on this topic, see e.g. Ref. 4.

Most of the studies reported above concern the one-dimensional (1D) periodic
structures, as the 1D case, being relatively simple, allows an analytical
treatment. The multi-dimensional cases (the 2D case as in our present study,
or even the 3D case) are much more difficult to be accessed analytically. The
majority of these studies in multi-dimensional case are numeric, as using
plane-wave expansion, or finite difference time domain (FDTD) schemes. These
studies also mostly concern the modification of the \textit{temporal}
dispersion characteristics.

It comes out recently, that the spatial periodicity can affect not only
temporal dispersion, but also the \textit{spatial} one, i.e. the dependence of
the longitudinal component of the propagation constant versus the transverse
component. These results (again predominantly numeric) lead to so called
management of spatial dispersion, i.e. to the management of diffraction
properties of narrow beams. This idea led to the prediction of the negative
diffraction of light beams in photonic crystals$^{5}$, of sound beams in sonic
crystals$^{6}$, and of coherent atomic ensembles in Bose-Einstein condensates
in periodic potentials$^{7}$. In particular it has been found recently that
between the normal diffraction and negative diffraction regimes a strong
reduction of the diffraction can be achieved, leading to the so called
self-collimating, or nondiffractive light beams$^{8} $.

The geometrical interpretation of wave diffraction is as follows: wave beams
of arbitrary shape can be Fourier decomposed into plane waves, which in
propagation acquire phase shifts depending on their propagation angles. This
dephasing of the plane wave components results in a diffractive broadening of
the beams. Fig.1(a) illustrates normal diffraction in propagation through an
homogeneous material, where the longitudinal component of the wavevector
depends trivially on the propagation angle, $k_{||}=k_{z}=\sqrt{\left\vert
\mathbf{k}\right\vert ^{2}-\left\vert \mathbf{k}_{\bot}\right\vert ^{2}}$,
where $\mathbf{k}_{\bot}=\left(  k_{x},k_{y}\right)  $. In general, the normal
or positive diffraction means that the surfaces of constant frequency are
concave in the wavevector domain $\mathbf{k}=(k_{x},k_{y},k_{z})$, as
illustrated in Fig.1(a). The negative diffraction, as illustrated in Fig.1(b),
geometrically means that the surfaces of constant frequency are convex in
wavevector domain. The intermediate case of the vanishing diffraction is
illustrated in Fig.1(c), where the zero diffraction is supposed to occur at a
particular point in the wavevector domain where the curvature of the surfaces
of constant frequency becomes exactly zero. Zero diffraction physically means
that beams of arbitrary width can propagate without diffractive broadening or,
equivalently, that arbitrary wave structures can propagate without diffractive
\textquotedblleft smearing\textquotedblright.

The present study concerns the nondiffractive propagation of sound in periodic
acoustic materials (sonic crystals). We found, by applying the plane-wave
expansion method, the existence of nondiffractive regimes similar to those in
optics, or to those expected from Fig.1(c). We check the nondiffractive
propagation by integrating the wave equations by means of the FDTD technique.
Moreover, we also present the analytical treatment of the problem, leading to
analytic relations, which among other are useful for the planning of the
corresponding experiment, and for designing the possible applications.

In Section II of the article the propagation of sound is analysed by plane
wave expansion, leading to the spatial dispersion curves, and in particular
resulting into the straight (nondiffractive) segments of the spatial
dispersion curves. In this way the nondiffractive propagation regimes are
predicted. In the next Section III the FDTD calculations are performed in the
predicted nondiffractive regimes, and the nonspreading propagation of narrow
beams is demonstrated. Section IV is devoted to the analytical study, to the
derivation of analytical relations between parameters for the nondiffractive
propagation. Last Section contains the concluding remarks, where the results
are summarized and also the minimal size of the beams propagating
nondiffractively is evaluated.

\section{Dispersion in sonic crystals}

The propagation of sonic waves is determined by the following linear system
equations
\begin{subequations}
\label{linsys}%
\begin{align}
\rho\frac{\partial\mathbf{v}}{\partial t}  & =-\nabla p,\\
\frac{\partial p}{\partial t}  & =-B\nabla\mathbf{v,}%
\end{align}
where $B(\mathbf{r})$ is the bulk modulus, $\rho(\mathbf{r})$ is the density
(both dependent in space), $p(\mathbf{r},t)$ is the pressure (which are scalar
fields), and $\mathbf{v}(\mathbf{r},t)$ is the velocity vector field.

We define the relative values of the bulk modulus $\bar{B}(\mathbf{r}%
)=B(\mathbf{r})/B_{h}$ and the density $\bar{\rho}(\mathbf{r})=\rho
(\mathbf{r})/\rho_{h}$, normalizing to the corresponding parameters in the
host medium. Then, eliminating the velocity field in Eqs. (\ref{linsys}), we
obtain a wave equation describing the propagation of sound in the
inhomogeneous medium,
\end{subequations}
\begin{equation}
\frac{1}{\bar{B}(\mathbf{r})}\frac{\partial^{2}p(\mathbf{r},\tau)}%
{\partial\tau^{2}}-\nabla\left(  \frac{1}{\bar{\rho}(\mathbf{r})}\nabla
p(\mathbf{r},\tau)\right)  =0.\label{waveqn}%
\end{equation}
where $\tau=c_{h}t$ is a normalized time, that makes the velocity of sound in
the host medium $c_{h}\,$\ equal to unity, where $c_{h}=\sqrt{B_{h}/\rho_{h}}$.

We consider sound beams with harmonic temporal dependence. Then, the steadily
oscillating solution has the form $p(\mathbf{r},t)=p(\mathbf{r})e^{i\omega
\tau}$, which substituted in Eq. (\ref{waveqn}) leads to the eigenvalue
equation
\begin{equation}
\frac{\omega^{2}}{\bar{B}(\mathbf{r})}p(\mathbf{r})+\nabla\left(  \frac
{1}{\bar{\rho}(\mathbf{r})}\nabla p(\mathbf{r})\right)  =0.\label{eigeq}%
\end{equation}

For the subsequent analysis we consider a concrete geometry, where acoustic
waves propagate in a two-dimensional medium, formed by an squared array of
solid cylinders, with axis along \textit{y} direction and radius $r_{0}$, in a
host fluid medium. The coordinate $\mathbf{r}$ in Eq. (\ref{eigeq}) depends
now on longitudinal (\textit{z}) and transverse (\textit{x}) directions, and
$\nabla=\left(  \partial/\partial x,\partial/\partial z\right)  .$

The lattice defined though the centers of cylinders is given by the set
$R=\left\{  \mathbf{R}=n_{1}\mathbf{a}_{1}+n_{2}\mathbf{a}_{2};\,\,n_{1}%
,n_{2}\in N\right\}  $ of two-dimensional lattice vectors $\mathbf{R}$ that
are generated by the primitive translations $\mathbf{a}_{1}$ and
$\mathbf{a}_{2}$. The corresponding reciprocal lattice is defined though
$G=\left\{  \mathbf{G}:\mathbf{G}\cdot\mathbf{R}=2\pi n;\,\,n\in N\right\}  .$

A possible way of solving Eq. (\ref{eigeq}) is by means of the plane wave
expansion (PWE) method, which converts the differential equation into an
infinite matrix eigenvalue problem, which is then truncated and solved
numerically. By solving this eigenvalue problem the frequencies corresponding
to each Bloch wave can be obtained, providing the dispersion relationship and
band structure of the periodic medium.

The bulk modulus and density are periodic functions with the periodicity of
the lattice, and therefore contain all the information of the phononic
crystal. This implies that the material parameters can be represented by their
Fourier expansions on the basis of the reciprocal lattice,%
\begin{equation}
\bar{\rho}(\mathbf{r})^{-1}=\sum\limits_{\mathbf{G}}\rho_{\mathbf{G}}%
^{-1}e^{i\mathbf{G\cdot r}},\label{rhoexpand}%
\end{equation}%
\begin{equation}
\bar{B}(\mathbf{r})^{-1}=\sum\limits_{\mathbf{G}}b_{\mathbf{G}}^{-1}%
e^{i\mathbf{G\cdot r}}.\label{Bexpand}%
\end{equation}

On the other hand, the solutions $p(\mathbf{r})$ of Eq. (\ref{eigeq}) must be
also periodic with the periodicity of the lattice (Bloch-Floquet theorem), and
can be expanded as%
\begin{equation}
p(\mathbf{r})=e^{i\mathbf{k\cdot r}}\sum\limits_{\mathbf{G}}p_{\mathbf{k}%
,\mathbf{G}}\,e^{i\mathbf{G\cdot r}},\label{pexpand}%
\end{equation}
where $\mathbf{k}$ is a two-dimensional Bloch vector restricted to the first
Brillouin zone, and $\mathbf{G}$ denotes a reciprocal lattice vector. For a
square lattice, $\mathbf{G}=(2\pi/a)(n_{1}\mathbf{e}_{1}+n_{2}\mathbf{e}_{2})$
with $n_{1}$ and $n_{2}$ integers and \textit{a} being the lattice constant
(minimal distance between the centers of neighbor cylinders).

The coefficients in expansions (\ref{rhoexpand}) and (\ref{Bexpand}) can be
obtained from the inverse Fourier transform. For the inverse of mass density
the coefficients result$^{9}$%
\begin{equation}
\rho_{\mathbf{G}}^{-1}=\frac{1}{a^{2}}\iint\limits_{UC}\frac{1}{\bar{\rho
}(\mathbf{r})}d\mathbf{r}=\frac{\rho_{h}}{\rho_{c}}f+(1-f),\ \ \ \ \text{for
\ }\mathbf{G}=0,\label{G0}%
\end{equation}
which represents the average value of the density, and%
\begin{equation}
\rho_{\mathbf{G}}^{-1}=\frac{1}{a^{2}}\iint\limits_{UC}\frac
{e^{i\mathbf{G\cdot r}}}{\bar{\rho}(\mathbf{r})}d\mathbf{r}=\left(  \frac
{\rho_{h}}{\rho_{c}}-1\right)  \,2f\,\frac{J_{1}\left(  \left\vert
\mathbf{G}\right\vert r_{0}\right)  }{\left\vert \mathbf{G}\right\vert r_{0}%
},\,\ \ \ \text{for \ }\mathbf{G}\neq0,\label{Gno0}%
\end{equation}
where the integration extends over the two-dimensional unit cell, $J_{1}(x)$
is the Bessel function of the first kind and $f=\pi(r_{0}/a)^{2}$ is the
filling fraction. Exactly the same expressions, follow for the coefficients of
bulk modulus $b_{\mathbf{G}}^{-1}$, since the expansion has an analogous form.

In terms of the coefficients of the previous expansions, Eq. (\ref{eigeq})
becomes%
\begin{equation}
\sum\limits_{\mathbf{G}^{\prime}}\left[  \omega^{2}b_{\mathbf{G}%
-\mathbf{G}^{\prime}}^{-1}-\rho_{\mathbf{G}-\mathbf{G}^{\prime}}^{-1}\left(
\mathbf{k}+\mathbf{G}\right)  \cdot\left(  \mathbf{k}+\mathbf{G}^{\prime
}\right)  \right]  \,p_{\mathbf{G}^{\prime}}=0.\label{matrix}%
\end{equation}

Equation (\ref{matrix}) has been numerically solved considering 361 plane
waves in the expansion. The number of plane waves has been chosen in order to
ensure the convergence. Figure 2 shows the band structure for a square lattice
of steel cylinders ($\rho_{c}=7.8\;10^{3}$ Kg{\ }m$^{-3}$, $B_{c}%
=160\,\,10^{9}$ N{\ }m$^{-2}$) immersed in water ($\rho_{h}=10^{3}$ Kg
m$^{-3}$, $B_{h}=2.2\,\,10^{9}$ N m$^{-2}$). The dimensionless (reduced)
frequency $\Omega=\omega a/2\pi c_{h}$ is plotted in terms of the
dimensionless wavenumber of Bloch vector $\mathbf{K}=\mathbf{k}a/2\pi$.

From the solutions of Eq. (\ref{matrix}) we can also compute the isofrequency
contours. In Fig. 3 the results for the first and second bands are shown. In
both cases, the curves show a transition from convex to concave at a
particular frequency. The isofrequency contours at the transition point
acquire, as shown in the figure, the form of squares with rounded corners.
Consequently, there exist locally flat segments of the curve, where, in other
words, the spreading of the beam will be counteracted by the crystal
anisotropy. Similarly as for photonic crystals in optics the nondiffractive
propagation occurs along the diagonals of squares in the first propagation
band, and along the principal vectors of the lattices in the second band. The
\textquotedblleft rounded nondiffractive square\textquotedblright\ is situated
around the corner of Brillouin zone (denoted by M) for the first band, and
around the centre of Brillouin zone (denoted by $\Gamma$) in the second band.

\section{Numerical results}

In order to prove the nondiffractive propagation of sound in the sonic
crystal, Eqs. (\ref{linsys}) have been numerically integrated using the Finite
Difference in Time Domain (FDTD) method. FDTD is nowadays a standard
technique$^{10}$ for solving partial differential equations by integrating in
time, and by approximating the spatial derivatives by finite differences. The
incident acoustic beam has been simulated by a plane vibrating surface
radiating a harmonic wave with variable frequency $\omega$, describing an
acoustic transducer with a diameter of 3 cm. The front face of the crystal is
located close to the source, where the wavefront is nearly plane. The medium
parameters were chosen to correspond to numerically evaluated zero diffraction
point (by inspecting the isofrequency curves) in the previous section. For the
first band [Fig.3(a)] the isofrequency curve becomes locally flat for
$\Omega\approx0.54$, which corresponds to a real frequency of $f=\Omega
c_{h}/a\approx154\,$KHz, and for an incidence along [1,1] direction. Under
these conditions, the nondiffractive propagation is predicted to occur. The
result of the numerical simulation for these parameters is shown in Fig. 4
(left column). As expected, the beam propagates through the crystal without a
visible divergence. For the second band, the theory predicts that the
frequency for nondiffractive propagation increases roughly by the factor of
$\sqrt{2}$ with respect to the first band, and then occurs at $f\approx217\,$KHz.

We note that whereas the beam in homogeneous media broadens sensibly over the
propagation distance, the same beam in the sonic crystal propagates without
sensible broadening over much longer distances. Diffractive broadening in
homogeneous medium is quantified by the Rayleigh distance $L_{d}=ka^{2}/2$,
where \textit{a} is the radius of the source, and corresponds to the distance
from the source after which the beam broadens in a factor of $\sqrt{2}$. For
the two nondiffractive frequencies evaluated above, the Rayleigh distances are
7.3 cm for the first case, and 10.3 cm for the second case.

\section{Analytics for nondiffracting beams}

The precise analysis of an arbitrary field structure inside the crystal can
only be performed by considering the field expansion into an infinite set of
modes of the crystal (as stated by the Bloch theorem). The form given by Eqs.
(\ref{rhoexpand})--(\ref{pexpand}) must be assumed, whose unknown amplitudes
can be numerically evaluated. This is the basics of the PWE method used in
Sec. II for evaluate the band structure and dispersion curves of the crystal.
However, it is possible to develop an analytical theory if we consider the
particular case of a nondiffractive beam, since this nondiffractive solution
appears mainly due to the coupling of three modes, the homogeneous one and the
next low order modes. This situation is illustrated in Fig.5, where the three
intersecting circles, corresponding to the spatial dispersion curves of the
three modes (those with wavenumbers $\mathbf{k}$, $\mathbf{k}+\mathbf{G}_{1}$
and $\mathbf{k}+\mathbf{G}_{2}$) give rise to the nondiffractive effect. Due
to the interaction between the different spatial modes the degenerancy at the
intersections of the spatial dispersion curves is lifted, and the flat areas
in the dispersion curve can develop. The radiation belonging to these
interacting modes is the most relevant for the deformation of the dispersion
curves and to the appearance of the flat segments, i.e. is responsible for the
nondiffractive propagation. Therefore the other modes are irrelevant in the
\textquotedblleft nondiffractive\textquotedblright\ area (shadowed region in
Fig. 5), and the Bloch expansion can be simplified as%
\begin{equation}
p(\mathbf{r})=\left(  p_{0}+p_{1}e^{i\mathbf{G}_{1}\cdot\mathbf{r}}%
+p_{2}e^{i\mathbf{G}_{2}\cdot\mathbf{r}}\right)  ,\label{3mode}%
\end{equation}
where $\mathbf{G}_{1}$ and $\mathbf{G}_{2}$ are the basis vectors of the
reciprocal space.

Note that since the nondiffractive beam is expected to be highly directive,
only $\mathbf{G}$ vectors directed to the same direction as the wavevector
$\mathbf{k}$ are relevant in the analysis. The material parameters (being real
functions) must be however expanded into at least five modes. In particular,
the inverses of density and bulk modulus will be assumed of the form
\begin{subequations}
\label{5modes}%
\begin{align}
\bar{\rho}(\mathbf{r})^{-1}  & =\left(  b_{0}+b_{+1}e^{i\mathbf{G}_{1}%
\cdot\mathbf{r}}+b_{+2}e^{i\mathbf{G}_{2}\cdot\mathbf{r}}+b_{-1}%
e^{-i\mathbf{G}_{1}\cdot\mathbf{r}}+b_{-2}e^{-i\mathbf{G}_{2}\cdot\mathbf{r}%
}\right)  ,\\
\bar{B}(\mathbf{r})^{-1}  & =\left(  b_{0}+b_{+1}e^{i\mathbf{G}_{1}%
\cdot\mathbf{r}}+b_{+2}e^{i\mathbf{G}_{2}\cdot\mathbf{r}}+b_{-1}%
e^{-i\mathbf{G}_{1}\cdot\mathbf{r}}+b_{-2}e^{-i\mathbf{G}_{2}\cdot\mathbf{r}%
}\right)  ,
\end{align}
where the notation $\rho_{\pm j}=\rho_{\pm\mathbf{G}_{j}}$, with
\textit{j}=1,2 has been used. Substituting Eqs. (\ref{5modes}) in Eq.
(\ref{eigeq}), and collecting the terms at the same exponents (those with
wavevectors $\mathbf{k}$, $\mathbf{k}+\mathbf{G}_{1}$ and $\mathbf{k}%
+\mathbf{G}_{2}$), we obtain the following coupled equation system,
\end{subequations}
\begin{subequations}
\label{eqsmodes}%
\begin{align}
0  & =\omega^{2}\left(  p_{0}b_{0}+p_{1}b_{-1}+p_{2}b_{-2}\right)
-\mathbf{k}^{2}p_{0}\rho_{0}-\mathbf{k}\left(  \mathbf{k}+\mathbf{G}%
_{1}\right)  p_{1}\rho_{-1}-\mathbf{k}\left(  \mathbf{k}+\mathbf{G}%
_{2}\right)  p_{2}\rho_{-2},\\
0  & =\omega^{2}\left(  p_{1}b_{0}+p_{0}b_{+1}\right)  -\left(  \mathbf{k}%
+\mathbf{G}_{1}\right)  ^{2}p_{1}\rho_{0}-\mathbf{k}\left(  \mathbf{k}%
+\mathbf{G}_{1}\right)  p_{0}\rho_{+1},\\
0  & =\omega^{2}\left(  p_{2}b_{0}+p_{0}b_{+2}\right)  -\left(  \mathbf{k}%
+\mathbf{G}_{2}\right)  ^{2}p_{2}\rho_{0}-\mathbf{k}\left(  \mathbf{k}%
+\mathbf{G}_{2}\right)  p_{0}\rho_{+2}.
\end{align}

Equations (\ref{eqsmodes}) are still too complex to lead to analytical
results. However, for small values of the filling fraction \textit{f} the
solid inclusions can be considered as a perturbation of the homogeneous fluid
medium, and an assymptotic analysis near the bandgap is justified. Next we
show that, in this limit, it is possible to obtain a simple relation between
the frequency and wavenumber of the nondiffractive beam and the filling
fraction \textit{f} characterizing the crystal.

First, note that in the case of small \textit{f} (i.e. when $r_{0}<<a$) and
materials with high-contrast, where $\rho_{h}<<\rho_{c}$ and $B_{h}<<B_{c}$
(as occurs, e.g., for steel cylinders in water), the coefficients of the
medium expansions in Eqs. (\ref{G0}) and (\ref{Gno0}) simplify to $\rho
_{0}=b_{0}=1-f$ and $\rho_{i}=b_{i}=-f$, for $i=\pm1,\pm2.$ Then, Eqs.
(\ref{eqsmodes}), expressed in matrix form, read
\end{subequations}
\begin{equation}
\left(
\begin{array}
[c]{ccc}%
\left(  1-f\right)  \left(  \omega^{2}-\mathbf{k}^{2}\right)  & f\left(
\mathbf{k}\left(  \mathbf{k}+\mathbf{G}_{1}\right)  -\omega^{2}\right)  &
f\left(  \mathbf{k}\left(  \mathbf{k}+\mathbf{G}_{2}\right)  -\omega
^{2}\right) \\
f\left(  \mathbf{k}\left(  \mathbf{k}+\mathbf{G}_{1}\right)  -\omega
^{2}\right)  & \left(  1-f\right)  \left(  \omega^{2}-\left(  \mathbf{k}%
+\mathbf{G}_{1}\right)  ^{2}\right)  & 0\\
f\left(  \mathbf{k}\left(  \mathbf{k}+\mathbf{G}_{2}\right)  -\omega
^{2}\right)  & 0 & \left(  1-f\right)  \left(  \omega^{2}-\left(
\mathbf{k}+\mathbf{G}_{2}\right)  ^{2}\right)
\end{array}
\right)  \left(
\begin{array}
[c]{c}%
p_{0}\\
p_{1}\\
p_{2}%
\end{array}
\right)  =0.\label{matrix2}%
\end{equation}

The aim is to obtain the values of $\omega$ and $\mathbf{k}$ for which the
beam propagates without diffraction. For a crystal with square symmetry, the
direction of the nondiffractive propagation with respect the crystal axes can
be obtained from the isofrequency curves evaluated in Sec. II. For the first
band (Figs. 3(a) and 4) nondiffractive propagation occurs for beams
propagating at 45$^{\circ}$ with respect to the crystal axes, i.e. in the
[1,1] direction. For our analysis, is convenient to consider the beam axis to
be coincident with the \textit{z} direction, and define a set of unitary basis
vectors as $\mathbf{G}_{1}=(-1,1)/\sqrt{2}$, $\mathbf{G}_{2}=(1,1)/\sqrt{2}$.
In this way, all magnitudes in reciprocal space are normalized by $\pi/a$.

For small \textit{f}, one also expects that the parameters corresponding to
the nondiffractive regime take values close to those in the bandgap (near the
corner of the first Brillouin zone; see Fig. 2). The wavenumber corresponding
to the first bandgap is then $\mathbf{K}_{B}=(0,1)/\sqrt{2}$ (remind that, in
normalized reciprocal space, $\mathbf{K}=\mathbf{k}a/2\pi$). In order to
investigate the behaviour of dispersion curves close to this point, we
consider waves with wavevector $\mathbf{K}=\mathbf{K}_{B}(1-\delta\mathbf{K})$
with $\delta\mathbf{K}=\left(  \delta K_{x},\delta K_{z}\right)  $
representing small deviations. We further assume that the frequency is close
to- (but less than) that corresponding to the bandgap, $\Omega=\Omega
_{B}(1-\delta\Omega)$, with the normalized gap frequency given by $\Omega
_{B}=1/\sqrt{2}$.

The solvability condition of Eq. (\ref{matrix2}) results from equating to zero
the determinant of the matrix, and leads to the relation in the form $F\left(
\delta\Omega,\delta K_{z},\delta K_{x};f\right)  =0$. Expanding for small
$\delta\mathbf{K}=\left(  \delta K_{z},\delta K_{x}\right)  $, an analytical
transversal dispersion relation $\delta K_{z}\left(  \delta K_{x}\right)  $ is
obtained, which allows to calculate the diffraction coefficient as the
curvature of the transverse dispersion curve, i.e., $D=(1/2)\partial^{2}\delta
K_{z}/\partial\delta K_{x}^{2}$. The nondiffractive point corresponds to
$D=0$. This expression is analytical but still cumbersome. However, assuming
that $f\sim\mathcal{O}(\varepsilon^{2}) $ and $\delta\Omega\sim\mathcal{O}%
(\varepsilon),$ where $\varepsilon$ is a smallness parameter, the following
simple analytical relation is obtained at the leading order:%
\begin{equation}
\delta\Omega_{ND}^{(1)}=f^{2/3}+\mathcal{O}\left(  f^{4/3}\right)
.\label{frecND}%
\end{equation}

Also the wavenumber of the nondiffractive beam can be analytically evaluated.
Substituting Eq. (\ref{frecND}) in the solvability condition of Eq.
(\ref{matrix2}), and assuming the above smallness conditions, it follows that,%
\begin{equation}
\delta K_{ND}^{(1)}=f^{2/3}-f^{4/3}+\frac{3}{4}f^{2}+\mathcal{O}%
(f^{7/3}).\label{KND}%
\end{equation}

For the second band, a similar analysis can be performed. The three-mode
expansion is illustrated in Fig. 6. In this case it is more convenient to use
the vector basis $\mathbf{G}_{1}=(1,0)$ and $\mathbf{G}_{2}=(0,1)$, and now
the nondiffractive beam propagation occurs in a direction coincident with one
of the lattice vectors. The parameters of the gap are in this case
$\mathbf{K}_{B}=(1,0)$ and $\Omega_{B}=1$. An assymptotic analysis similar to
that performed above for the first band, shows that $\delta\Omega_{ND}%
^{(2)}=\delta\Omega_{ND}^{(1)}$ and $\delta\mathbf{K}_{ND}^{(2)}%
=\,\delta\mathbf{K}_{ND}^{(1)}$. Then, from the analytics follows that, under
the limit of the weak modulation that $f\sim\mathcal{O}(\varepsilon^{2})$ and
$\delta\Omega\sim\mathcal{O}(\varepsilon)$, the zero diffraction point for
both bands are situated equally, however with respect to the corresponding
bandgap: the diffraction in the first band disappears just below the first
bangap (by $\delta\Omega_{ND}=f^{2/3}$), and in the second band just below the
second bandgap by the same value (by $\delta\Omega_{ND}$). The wavevector
shifts are also equal for both cases. As a consequence, Eqs. (\ref{frecND})
and (\ref{KND}) are valid for both bands.

These analytical predictions have been checked numerically. In Fig.7 the
analytical results given by Eqs. (\ref{frecND}) and (\ref{KND}) are compared
with the corresponding numerical results (with symbols) obtained with the PWE
method using 361 modes. The curve labeled (a) corresponds to the normalized
frequency shift, measured with respect to the bandgap, and the curve labeled
(b) to the wavenumber shift, for zero diffraction point in the first band
(circles) and the second band (squares). The simple expressions obtained under
the three-mode expansions are in very with good agreement with the numerical
results, even for the moderate (not very small) values of the filling factor
\textit{f}.

\section{Conclusions}

Concluding, we have demonstrated theoretically the possibility of
nondiffractive propagation of acoustic beams through sonic crystals. We show
the nondiffractive propagation for both propagation bands: for the first band,
where the nondiffractive propagation occurs along the diagonals of the
lattice, and for the second band, where diffraction disappears along the
principal vectors of the lattice. The diffraction disappears for frequencies
just below the corresponding bandgaps, with equal frequency shift for both
cases given by a universal and very simple expression: $\delta\Omega
_{ND}=f^{2/3}$.

The other universal relation (\ref{KND}), for the shift of the wavenumber,
which in simplified form reads $\delta k_{ND}=f^{2/3}$, allows to evaluate the
width of the nondiffractively propagating beams. Indeed the half-width of the
platoe of spatial dispersion curve is roughly equal to (slightly less than)
$\delta k_{ND}$. This means that beams with normalized size $d\approx
2\pi/\delta k_{ND}\approx2\pi\,f^{-2/3}$ can propagate nondiffractively over
large distances (comparing with the diffraction length of the corresponding
beam in the homogeneous materials). In the terms of non-normalized
coordinates, the minimum size of the beam corresponds to $d\approx\sqrt
{2}a\,f^{-2/3}$. For a filling factor $f=0.114$, corresponding to the
numerical simulation in Fig. 4, the width of the nondiffractive beam predicted
by this expression results $d\approx6a$, i.e. to 6 spatial periods of the
lattice. This result is in good agreement with the width of the beam observed
in Fig. 4.

\medskip

{\large Acknowledgements}\medskip

The work has been partially supported by project FIS2005-07931-C03-01, and -03
of the Spanish Ministry of Science and Technology.

\newpage\bigskip{\large FIGURE CAPTIONS}\bigskip

\textbf{Fig.1.} Geometrical interpretation of diffraction of waves propagating
along the $z$ axis: a) positive, or normal diffraction in propagation through
homogeneous materials; b) negative, or anomalous diffraction; c) zero
diffraction. The area of negligible diffraction (for evaluation of the minimum
size of the nondiffractive beam) is indicated.

\textbf{Fig. 2.} Band structure for steel cylinders in water, for $r=1$\ mm,
$a=5.25$ mm, as calculated by the expansion into Bloch modes (4)-(7). The
solid squares mark the nondiffractive points (see Fig.3).

\textbf{Fig.3} Isofrequency lines, evaluated for $a=5.25$ mm and $r=1$ mm, for
the first (a) and second (b) bands, centered at $\Gamma$ point, as calculated
by the expansion into Bloch modes (4)-(7).

\textbf{Fig.4}. Numerical FDTD simulation of the nondiffractive beam, for the
first two bands. Upper row corresponds to the field radiated by the source
without crystal, lower row to the nondiffractive propagation in the [1,1]
(left) and [1,0] (right) directions, at the frequencies equal to $f=154\,$KHz
and $f=217\,$KHz as predicted by the theory. Gray levels are in decibel scale,
and the coordinates in meters. The other parameters are as in Figs. 2 and 3,
i.e. steel cylinders in water are simulated with $r=1$ mm and $a=5.25$ mm.

\textbf{Fig. 5.} Schematic picture showing the nondiffractive region (shadowed
area) resulting from the interaction of three modes. The square represents the
limits of the first Brillouin zone. The insert illustrates the lift of the
degenerancy at the cross sections of the dispersion curves and the formation
of the Bloch modes. The upper Bloch mode can develop flat segments depending
on the interaction strength, as the degree of the lift of degenerancy is
proportional to the interaction strength.

\textbf{Fig.6.} Schematic picture showing the nondiffractive region (shadowed
area) in the second propagation band. Everything as in Fig.5. Here the most
relevant modes are $\mathbf{k}+\mathbf{G}_{1}$ and $\mathbf{k}\pm
\mathbf{G}_{2}.$

\textbf{Fig.7.} Dependence of the frequency (a) and wavenumber of
nondiffractive beam, measured with respect to the bandgap values, as results
from numerical calculation (symbols) and the analytical expressions given in
Eqs. (12) and (13). The open circles and squares correspond to the parameter
values used for FDTD calculation of nondiffractive propagation (Fig.4) in the
first and second band respectively.


\begin{thebibliography}{99}                                                                                               %
\bibitem {1}E.Yablonovitch, \textit{Inhibited Spontaneous Emission in
Solid-State Physics and Electronics}, Phys. Rev. Lett. \textbf{58}, 2059
(1987); S. John, \textit{Strong localization of photons in certain disordered
dielectric superlattices}\textbf{,} Phys. Rev. Lett. \textbf{58}, 2486 (1987).

\bibitem {2}See e.g. \textit{Photonic Band Gaps and Localization}, edited by
C.M.Soukoulis NATO Advanced Studies Institute, Series B: Physics, Vol.308
(Plenum, New York, 1993).

\bibitem {3}M. Scalora, R. J. Flynn, S. B. Reinhardt, R. L. Fork, M. J.
Bloemer, M. D. Tocci, C. M. Bowden, H. S. Ledbetter, J. M. Bendickson, J. P.
Dowling and R. P. Leavitt.\textit{,} \textit{Ultrashort pulse propagation at
the photonic band edge: Large tunable group delay with minimal distortion and
loss}\textit{,} Phys. Rev. E \textbf{54}, R1078 (1996), Imhof A., Vos WL,
Sprik R, and Lagendijk A, \textit{Large Dispersive Effects near the Band Edges
of Photonic Crystals}\textit{,} Phys. Rev. Lett. \textbf{83}, 2942 (1999); K.
Sakoda, \textit{Enhanced light amplification due to group-velocity anomaly
peculiar to two- and three-dimensional photonic crystals}, Opt. Express
\textbf{4} 167 (1999).

\bibitem {4}T. Miyashita, \textit{Sonic crystals and sonic wave-guides}, Meas.
Sci. Technol. 16, R47-R63 (2005); Page JH, Sukhovich A, Yang S, Cowan ML, Van
der Viest F, Tourin A, Fink M, Liu Z, Chan CT and Sheng P, \textit{Phononic
crystals}, Phys. Stat. Sol. (b) \textbf{241}, 3454-3462 (2004)

\bibitem {5}R. Morandotti, H. S. Eisenberg, Y. Silberberg, M. Sorel and J. S.
Aitchison, \textit{Self-Focusing and Defocusing in Waveguide Arrays}\textbf{,}
Phys. Rev. Lett. \textbf{86}, 3296 (2001), M.J. Ablowitz and Z.H. Musslimani,
\textit{Discrete Diffraction Managed Spatial Solitons}\textbf{,} Phys. Rev.
Lett. \textbf{87}, 254102 (2001).

\bibitem {6}Suxia Yang, Page JH, Liu Z, Cowan ML, Chang CT and Sheng P,
\textit{Focusing of Sound in a 3D Phononic Crystal}, Phys. Rev. Lett.
\textbf{93}, 024301 (2004); M. Torres and F.R. Montero de Espinosa,
Ultrasonics \textbf{42}, 787 (2004)

\bibitem {7}E.A.Ostrovskaya and Yu.S.Kivshar, \textit{Matter-Wave Gap Solitons
in Atomic Band-Gap Structures}, Phys. Rev. Lett. \textbf{90}, 160407 (2003);
C.Conti and S.Trillo, \textit{Nonspreading Wave Packets in Three Dimensions
Formed by an Ultracold Bose Gas in an Optical Lattice}, Phys. Rev. Lett.
\textbf{92}, 120404 (2004).

\bibitem {8}Kosaka H, Kawashima T, Tomita A, Notomi M, Tamamura T, Sato T and
Kawakami S, \textit{Self-collimating phenomena in photonic crystals}, Appl.
Phys. Lett. \textbf{74}, 1212 (1999); Chigrin D, Enoch S, Sotomayor Torres C
and Tayeb G, \textit{Self-guiding in two-dimensional photonic crystals},
Optics Express, \textbf{11}, 1203 (2003); R. Iliew, C. Etrich, U. Peschel, F.
Lederer, M. Augustin, H.-J. Fuchs, D. Schelle, E.-B. Kley, S. Nolte, and A.
T\"{u}nnermann, \textit{Diffractionless propagation of light in a low-index
photonic-crystal film}, Appl. Phys. Lett. \textbf{85}, 5854 (2004);{\ }D. W.
Prather, S. Shi, D. M. Pustai, C. Chen, S. Venkataraman, A. Sharkawy, G.
Schneider, and J. Murakowski, \textit{Dispersion-based optical routing in
photonic crystals}, Opt. Lett. \textbf{29}, 50-52 (2004); K.Staliunas and R.
Herrero, \textit{Nondiffractive propagation of light in photonic crystals},
Phys. Rev. E, \textbf{73}, 016601 (2006)

\bibitem {9}M.S. Kushwaha and P. Halevi, \textit{Gian stop bands in
two-dimensional periodic arrays of liquid cyslinders,} Appl. Phys. Lett.
\textbf{69}, 31 (1996)

\bibitem {10}Vasseur J O, Djafari-Rouhani B, Dobrzy'nski L, Kushwaha M M and
Halevi P, J. Phys.:Condens. Matter \textbf{6}, 8759 (1994).
\end{thebibliography}
\end{document}